\documentclass[aps,prl,twocolumn,showpacs,preprintnumbers,amsmath,amssymb,superscriptaddress]{revtex4}%
\usepackage{epsfig}%
\usepackage{dcolumn}
\usepackage{bm}
\begin{document}%
\title{ 
The microscopic study of a single hydrogen-like impurity 
in semi-insulating GaAs
}%
\author{D.G.~Eshchenko}
\affiliation{Physik-Institut der Universit\"{a}t Z\"{u}rich,
Winterthurerstrasse 190, CH-8057, Z\"urich, Switzerland}
\affiliation{ Laboratory for Muon Spin Spectroscopy, 
Paul Scherrer Institut, CH-5232 Villigen PSI, Switzerland}
\email{dimitry.eshchenko@psi.ch}
\author{V.G.~Storchak}
\affiliation{Russian Research Centre ``Kurchatov Institute",
 Kurchatov Sq.~46, Moscow 123182, Russia}
\author{S.P.~Cottrell}
\affiliation{ISIS Facility, Rutherford Appleton Laboratory,
Oxfordshire OX11 OQX, UK}
\author{E.~Morenzoni}
\affiliation{ Laboratory for Muon Spin Spectroscopy, 
Paul Scherrer Institut, CH-5232 Villigen PSI, Switzerland}

\begin{abstract}
The charge dynamics of hydrogen-like centers formed by the implantation of
energetic (4~MeV) muons in semi-insulating GaAs have been studied by muon
spin resonance in electric fields. The results point to the significant role of
deep hole traps in the compensation mechanism of GaAs. Electric-field-enhanced
neutralization of deep electron and hole traps by muon-track-induced hot 
carriers results to an increase of the non-equilibrium carrier life-times. As
a consequence, the muonium ($\mu^+ + e^-$) center 
%
at the tetrahedral As site can capture the track's holes and therefore behaves like a donor.

\end{abstract}
\vfil

\maketitle%

Semi-insulating (SI) GaAs is an important material that
forms the basis of the GaAs microwave and integrated circuit industries. 
SI substrates are used for the growing of GaMnAs thin films, recognised as a 
prototype material for future 
spintronics devices \cite{Ohno_1998}.
The semi-insulating properties of GaAs arise from the compensation 
of residual shallow donors, residual or intentionally added shallow acceptors 
and intrinsic or intentionally introduced deep centers.  
Deep centers may act as carrier traps, recombination
centers or scattering centers, and have a strong influence
on the electronic properties of the material even
when their concentration is much less than the carrier
density.     
Hence studies of deep level defects in GaAs is of both fundamental and
applied interest.

Since the passivation of shallow acceptors and donors in GaAs by hydrogen was 
discovered in 1986 \cite{Johnson_1986}, the behaviour of 
hydrogen in III­-V semiconductors 
has 
became the subject of intense research. 
Significant information 
on the microscopic properties of
isolated hydrogen impurities 
have been obtained 
from muon spin
rotation spectroscopy, where the behaviour of positive muons and muonium (Mu$ = \mu^+ +
e^-$) are studied as an analogue of a light hydrogen isotope (m$_\mu \simeq 1/9m_{\rm p}$). 
In this letter we demonstrate that $\mu$SR can be used to get 
alternative information
on deep 
centers in GaAs.

Muonium 
in GaAs 
can exist in three charge states Mu$^+$, Mu$^0$ and Mu$^-$
\cite{Patterson,Lichti_2007}. 
The stable position for Mu$^+$ lies on the Ga-As bond (so-called Mu$^+_{\rm
BC}$), with another possible location being the anti-bonding position close to As atom 
(AB$_{\rm As}$). The tetrahedral void  
between four As atoms  (T$_{\rm As}$) is also considered \cite{Lichti_2007} as 
a metastable
position supporting Mu$^+$. Negative Mu$^-$ is usially placed at the tetrahedral 
position formed by four Ga atoms Mu$^-_{\rm T_{Ga}}$, or at the 
anti-bonding position close to Ga atom (AB$_{\rm Ga}$). 
Neutral 
muonium 
is 
believed to occupy 
the bond-center position (Mu$^0_{\rm BC}$) and 
both T$_{\rm Ga}$ and T$_{\rm As}$ tetrahedral sites and is thus denoted by 
Mu$^0_{\rm T}$.

In semi-insulating and slightly doped
n-type GaAs, 
the following is the accepted picture describing the 
distribution of muonium between the different states on a 
nanosecond  
time-scale following muon implantation. At low
temperatures (T$<100$~K) the diamagnetic fraction is 
small, while the Mu$^0_{\rm BC}$ and Mu$^0_{\rm T}$
fractions are about 40(5)\% and 60(5)\% respectively. The 
Mu$^0_{\rm T}$ fraction remains constant up to at least room temperature, while 
the Mu$^0_{\rm BC}$ fraction decreases above 100~K. This process is accompanied  
by a corresponding increase in
the diamagnetic fraction, and is usually described in terms of Mu$^0_{\rm BC}$ 
to Mu$^+_{\rm BC}$ ionization \cite{Patterson} (implying the thermal emission 
of the electron from the Mu$^0_{\rm BC}$ to the conduction band). At low
temperatures, electric field experiments 
\cite{Eshchenko_1999,Eshchenko_Cr_2002} have shown a gradual suppression of the 
Mu$^0_{\rm BC}$ signal by the application of an electric field of the order $E_{\rm char} \sim 5$~kV/cm. The reduction of the Mu$^0_{\rm BC}$
fraction is accompanied by a corresponding 
increase in the diamagnetic signal; however,
electric fields of up to 20~kV/cm have no effect on the Mu$^0_{\rm T}$ signal. 
In this letter we show that very small electric field of $\sim 1-2$~kV/cm can
effect long term ($\sim 10^{-6}$~s) 
dynamics of the neutral muonium. 

Most of the experiments reported here were performed at the ISIS pulsed muon facility, located
at the Rutherford Appleton Laboratory (RAL, Chilton, UK), using the novel technique of
radio-frequency (RF)-$\mu$SR in electric fields 
EF-RF-$\mu$SR\cite{Eshchenko_RF}. 
RF experiments
are essentially longitudinal magnetic field ($LF$) measurements, where the
initial 
direction of the muon spin is co-linear with the relatively large 
(in our case about $1.5-2$~kOe)
magnetic field, with
a small ($H1\sim 10$~Oe) RF field applied perpendicular to the   
$LF$ at a frequency tuned to match the Zeeman splitting   
of the diamagnetic species ($\sim 20$~MHz).
Diamagnetic states formed from a paramagnetic precursor, even on a microsecond timescale, will contribute to the precessing RF asymmetry asymmetry, and consequently RF techniques have become a well established tool 
for studying the dynamics of slow muonium to diamagnetic conversion
\cite{Morozumi_1986}.
Our previous EF-$\mu$SR experiments in insulators \cite{Eshchenko_2002} demonstrated that the muon is surrounded by its own track products, with the main
radiation damage (the majority of electron-holes pairs) being behind the stopped
muon.  In the flat geometry we are using,  
by choosing the polarity of the applied  electric field
(parallel or anti-parallel to the muon track direction) one can select 
the appropriate charge (electrons or holes)  for the long range interactions with
the stopped muon.    
Thus EF-RF-$\mu$SR \cite{Eshchenko_RF} 
enables the selective 
(electrons vs holes) study of the  
dynamics of the interaction of muonium with hot carriers on a microsecond timescale.    
Since the stopped muon is a single probe at the end of his own
track, the dynamics of track products will be determined by the intrinsic properties
of the sample under the study e.g by 
deep centers.
In this respect, $\mu$SR experiments may offer a unique insight on the
fundamental non-equilibrium properties of 
deep centers. 

For $\mu$SR measurements, 100\% spin-polarized positive muons
are implanted into the sample and
the positron decay products
monitored. Due to parity violation, the positrons are emitted
preferentially along the instantaneous direction of the muon
spin. Using two sets of positron counters placed upstream and downstream of the sample
(backward and forward counters) the time-differential asymmetry of the muon 
decay is constructed
in the standard way as $A(t)=\frac{N_B(t)-N_F(t)}{N_F(t)+N_B(t)}$, where $N_B(t)$
and $N_F(t)$ are decay counts in the backward and forward detectors
respectively. 
The maximum asymmetry for the apparatus depends on the positions
and solid angles of the
detectors and, in our case, is about 0.23. Each $\mu$SR spectrum contains at
least several million decays.  
Four sets of histograms are collected: 
RF field on, positive
electric field; RF field on, negative electric field; RF off, positive electric
field; RF off, negative electric field. The positive electric field points 
parallel to the initial muon momentum. The states were changed every 1/50~s at every ISIS accelerator frame (i.e. before every muon pulse) 
\begin{figure}[h]
\begin{center}
\epsfig{file=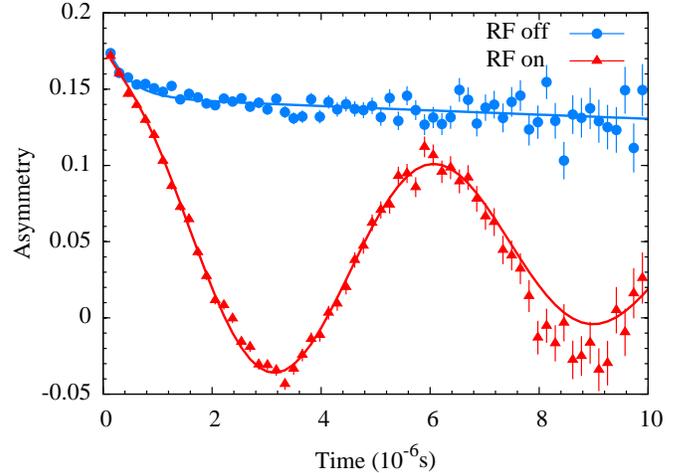,width=\columnwidth}
\end{center}
\vspace{-0.5cm}
\caption{Typical experimental spectra measured in semi-insulating GaAs 
after the electric field treatment
measured at $T=70$~K in longitudinal
magnetic field $LF=1837$~Oe and zero electric field. Blue circles - RF signal is
off; red triangles - RF signal is on. 
}
\label{70K}
\end{figure}
Typical $\mu$SR spectra measured in GaAs are presented in Fig.~(\ref{70K}).
The RF off asymmetry spectrum consists of a fast-relaxing Mu$_{\rm T}$ fraction and
a slow-relaxing signal which is the sum of the Mu$^0_{\rm BC}$ and
diamagnetic components. The contributions of these two components are resolved 
in the RF spectrum, where the diamagnetic fraction is seen in the precessing signal 
(which
has a slow relaxation due mainly to the space inhomogeneity of the RF field) 
while the
non-relaxing Mu$^0_{\rm BC}$ 
is observed as a constant offset in the spectra.

Studies reported here were performed using a commercial 
high resistivity GaAs substrate
($\rho \sim (1.6-1.8)\times10^8$~Ohm$\times$cm, n-type
conductivity with an electron mobility $\sim 5000$~cm$^2$/V$\cdot$s), 
 0.5~mm thick, with the $<100>$ crystallographic axis perpendicular 
to the surface impacted by the muon beam. The sample was purchased 
from the American Xtal company, where it was grown using the vertical boat (VB) technique. 
Silver electrodes of $\sim$80~nm thickness were deposited on both
surfaces of the sample using the DC magnetron sputtering technique, thereby making two Schottky contacts. This allowed us to apply 
voltages of up to 1~kV in both polarities, giving us a nominal electric field in the sample of up to $\pm$20~kV/cm. 

\begin{figure}[h]
\begin{center}
\epsfig{file=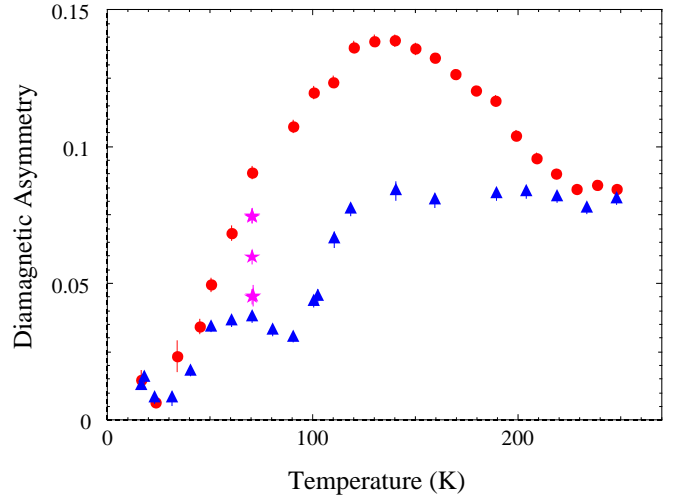,width=\columnwidth}
\end{center}
\vspace{-0.5cm}
\caption{Temperature dependencies of the diamagnetic signals measured at zero
electric field in
different states of commercial si-GaAs.  Triangles: RF diamagnetic
asymmetry in the virgin sample. Circles:  enhanced RF signal in the treated
sample. Stars: RF asymmetry measured in the process of the electric field
treatment.   
}
\label{RF}
\end{figure}

%
%
A measurement of 
the  
diamagnetic fraction 
in zero electric 
field is shown in Fig.~(\ref{RF}).  
The triangles represent the temperature dependence of the RF diamagnetic asymmetry in
the virgin sample (no electric field history). This dependence is similar to
that reported in the literature \cite{Hitti_2000,Lichti_2007}. 
The 
increase in
the asymmetry 
is described in terms of
Mu$^0_{\rm BC}$ to Mu$^+_{\rm BC}$ conversion. Above 120~K the RF asymmetry
remains constant, and is in good agreement with the Mu$^0_{\rm BC}$ fraction measured
in transverse magnetic field experiments at low temperatures \cite{Patterson}. 
This implies that in this sample below 250~K there is no
contribution to the diamagnetic fraction from conversion of the Mu$^0_{\rm T}$ state, a conclusion in agreement with \cite{Lichti_2007}. 
The 
picture changes drastically if, at low temperatures, the sample is 
treated by applying a large ($E>10$~kV/cm) electric field simultaneously with 
muon implantation. The enhanced diamagnetic RF asymmetry is shown by the circles. Above $\sim$70~K, the diamagnetic asymmetry in the treated sample  exceeds that of 
the high temperature fraction measured in the virgin sample, implying an additional  channel for  
Mu$^0_{\rm T}$ muonium to a diamagnetic state conversion is opened by the sample treatment.
Measurements were carried out at a fixed temperature of 70~K to evaluate how the sample changes from its virgin condition to the treated state as a function of exposure to the muon beam. Results are shown by the stars in Fig.~(\ref{70K}), each corresponding to a zero electric field measurement carried out after a further 30 minutes of treatment in a field $E=\pm 16$~kV/cm. To complete the transition, about two hours of beam exposure or $\sim 1.5-3.0\times10^{15}$ electron-hole pairs per cubic centimeter are required.
The enhanced RF asymmetry passes through a broad maximum around 120--140~K (a state that persists for at least 12 hours), while
at temperatures above 140~K it shows annealing behavior and on warming to $\sim$220~K the sample is returned to the virgin state.

\begin{figure}[h]
\begin{center}
\epsfig{file=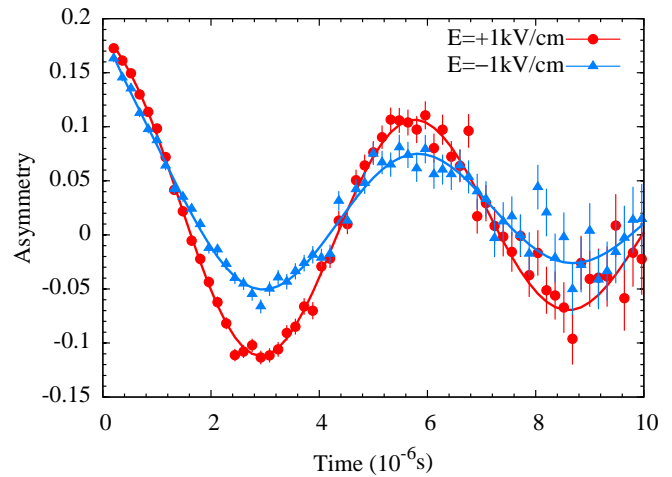,width=\columnwidth}
\end{center}
\vspace{-0.5cm}
\caption{Experimental spectra measured at $T=120$~K in SI-GaAs 
after 
electric field treatment. 
Red circles: $E=+1$~kV/cm. Blue triangles: $E=-1$~kV/cm.  Small negative
electric fields reduce the precessing RF amplitude 
to the value measured for the virgin
sample, while small positive electric fields have no effect on the amplitude. 
}
\label{low_field}
\end{figure}

The origin of the enhanced diamagnetic RF signal can be understood from the
electric field measurements.
Typical $\mu$SR spectra measured in the treated sample 
in electric fields $\pm$1~kV/cm 
are presented in  Fig.~(\ref{low_field}).
Small negative 
electric fields reduce the precessing RF amplitude 
to the value measured for the virgin
sample, while small positive electric fields have no effect on the amplitude. 
Taking into account the direction of the effect (positive electric fields point
parallel to the incoming muon beam) and recognizing that the injected
track carriers are positioned around and mostly behind the stopped muon, 
the reason for the enhanced diamagnetic signal seems likely to arise from the interaction between the track holes and the 
Mu$^0_{\rm T_{As}}$ species. A negative electric field will
pull holes away from the stopped muon, thereby reducing the muonium to diamagnetic
conversion, while a positive electric field will pull out the nearest holes but 
push "early"
track holes towards the stopped muon and muonium to to diamagnetic
conversion is unchanged. 
%
%
%
In the virgin sample, a small electric field does not change the RF precessing
asymmetry.

\begin{figure}[h]
\begin{center}
\epsfig{file=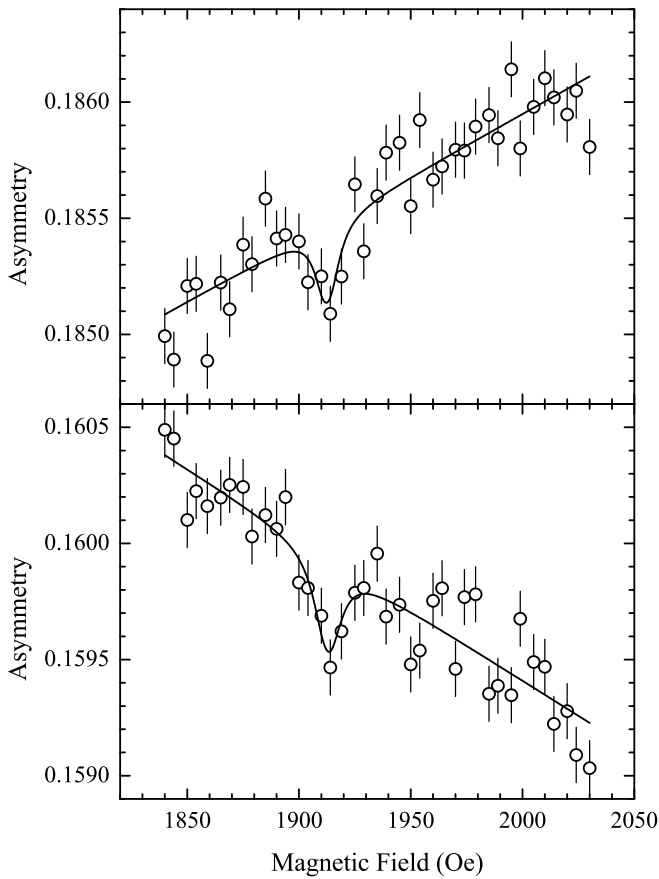,width=\columnwidth}
\end{center}
\vspace{-0.5cm}
\caption{Muon-nuclear level crossing integral spectra measured at $T=120$~K 
in SI-GaAs 
in the treated state. 
Top panel: $E=+1.5$~kV/cm. Bottom panel: $E=-1.5$~kV/cm. Electric field is
switched with $f=1$~kHz. The amplitude of the
resonance at $\sim$1913~Oe 
does not depend on the direction of a small electric field. 
}
\label{ALC}
\end{figure}

To identify the lattice position of the 
diamanetic species involved in the enhanced state,
muon-nuclear level crossing measurements
$\mu$LCR experiments, were perfomed 
at the Paul Scherrer 
Institute, Switzerland. 
For Mu$^+_{\rm BC}$ the main resonance in our 
geometry (H parallel to $<100>$) is expected around  
$H\sim$1920~Oe \cite{Chow_2005}; 
for Mu$^-_{\rm T_{Ga}}$ the resonances are observed at $H<500$~Oe
\cite{Chow_1995}.
For a treated sample we failed to record any resonance 
in the field range of Mu$^-_{\rm T_{Ga}}$.  
Resonances were, however, detected
at a field of $\sim$1913~Oe 
for treated SI-GaAs at $T=120$~K, and results for
both small positive and negative electric fields are presented in Fig.~(\ref{ALC}). The resonance 
amplitude does not depend on the direction of the electric field, suggesting that 
the enhanced diamagnetic signal does not belong to the Mu$^+_{\rm BC}$ 
state 
(at 
least  
during the time window of $\mu$SR measurements $\sim 10^{-5}$~s).  
We therefore need to consider that the tetrahedral As void may be supporting the
Mu$^+_{\rm T_{As}}$ state. 

The semi-insulating properties of the commercial vertical boat (VB) GaAs substrates 
are achieved by the compensation of shallow (residual or intentionally added) 
centers by the deep native centers. 
There are two main models discribing the compensation mechanism.
A 
widely accepted theory \cite{Holmes_1982} suggests that the 
compensation can be 
attributed to the interaction 
between the deep intrinsic 
donors EL2 (an As$_{\rm Ga}$-related defect
which is present at the level $N_{\rm EL2} \sim 10^{16}$cm$^{-3}$), 
residual donors (usually Si) and shallow carbon acceptors which are 
intentionally 
introduced in GaAs at the level of 
$n_{\rm C} \sim 0.5\times10^{15}$cm$^{-3}$.
However, the
Fermi level (EF) is often found to lie below the EL2 midgap
level, implying the existence of deep acceptor defects which can act like
hole traps. 
Thermally stimulated (TS) current in conjunction
with TS Hall effect measurements \cite{Bourgoin_1996} have demonstrated that 
the concentration of the deep hole traps in SI-GaAs is  
of the same order 
as the concentration of the deep electron traps, and that 
the pinning of the Fermi level on the
EL2 midgap may be explained as a result of the balance between deep
electron and hole traps.

The capture of electrons by the EL2$^+$
traps occurs over a configuration barrier with energy of $\sim 0.066$~eV
\cite{Martin_1979}. 
The applied electric field increases the energy of the 
carriers
and results in  
enhanced phonon-assisted trapping over the configurational barrier.
A similar behavior is observed for the holes traps.
Capacitance spectroscopy measurements \cite{Prinz_1983} have 
revealed that at low temperatures ($T \simeq 90$~K) a strong electric field 
($E\simeq10^4$~V/cm) causes 
an increase of both electron and hole capture cross sections by five orders of magnitude for the deep
traps in GaAs. 

Based on these conclusions, the following model for the formation of the enhanced diamagnetic state measured in SI-GaAs is proposed. 
At room temperature, the initial state of the sample is characterized by ionized residual shallow donors, partially ionized EL2 traps
and filled
by electrons or "ionized" deep acceptor
levels (empty holes traps) and ionized shallow acceptor centers.
Muon implantation at low temperatures results in the production of 
non-eqilibrium track carriers (electrons and holes) inside the 
sample. If a large electric field is applied, 
these carriers are transported through the bulk of the sample and are effectively captured 
by deep traps. 
At low temperatures the thermal emission of 
electrons/holes from deep traps is negligible, and 
consequently the sample is 
transfered to a metastable-like state. 
If the traps 
are filled, the 
life-times of non-equilibrium carriers 
are increased and the 
track electrons/holes can interact with the neutral Mu$^0_{\rm T}$ on a
microsecond time scale.            
By considering how small electric fields of opposite polarity affect muonium to diamagnetic
state 
conversion one can conclude that muonium interacts with the its own track holes. 
As the temperature in increased, electrons and holes are emitted from the filled 
deep traps, life-times for the excess carriers are reduced and there are no
holes available to promote state conversion regardless of the
sign of the applied electric field. 
The essential part of this model is the presence of both types of 
deep traps. In contrast, if one follows the simplest model of the SI-GaAs compensation
mechanism (where only one deep EL2 electron trap is involved) then the electric
field treatment will result to the filling of these traps  
and the excess hole life-time would not increase.

In conclusion, we have used the $\mu$SR technique in combination with electric fields and RF measurements to probe 
non-equilibrium carrier dynamics in 
SI-GaAs. Below 120~K, the simultaneous bipolar
injection (from the muon beam) and treatment of the sample with high electric fields results 
to the neutralization of both electron and hole deep traps. In the neutralized 
sample, excess holes from the muon track can be captured by muonium specied formed in the void created by the tetrahedral cage of four As atoms,
where it acts as a filled donor. Our results 
point to the scenario where not only deep EL2 centers but also deep hole traps 
are involved in the compensation mechanism in SI-GaAs.     

This work was partially performed at ISIS Pulsed Muon Source, United Kingdom,  
and at the Swiss Muon Source Facility of the Paul Scherrer Institute, 
Villigen, Switzerland.


\begin{thebibliography}{99}

\bibitem{Ohno_1998}
H.~Ohno, 
Science {\bf 281}, 951 (1998).

\bibitem{Johnson_1986}
N.M.~Johnson, R.D.~Burnham, R.A.~Street, and R.L.~Thornton, 
Phys. Rev.~B {\bf 33}, 1102 (1986).

\bibitem{Patterson}
B.D~Patterson,
Rev. Mod. Phys. {\bf 60}, 69 (1988).

\bibitem{Lichti_2007}
R.L.~Lichti, H.N.~Bani-Salameh, B.R.~Carroll, K.H.~Chow, 
B.~Hitti and S.R.~Kreitzman,
Phys. Rev.~B {\bf 76}, 045221 (2007).

\bibitem{Eshchenko_1999}
D.G.~Eshchenko, V.G.~Storchak , and G.D. Morris, 
Phys. Lett.~A {\bf 264}, 226 (1999).

\bibitem{Eshchenko_Cr_2002}
D.G.~Eshchenko, V.G.~Storchak, J.H.~Brewer, R.L.~Lichti, 
Phys. Rev. Lett. {\bf 89}, 226601 (2002).

\bibitem{Eshchenko_RF}
D.G.~Eshchenko, V.G.~Storchak, B.~Hitti, S.R.~Kreitzman, 
J.H.~Brewer, K.H.~Chow, 
Physica B {\bf 326}, 244 (2003).


\bibitem{Morozumi_1986}
Y.~Morozumi, K.~Nishiyama, and K.~Nagamine,
Phys. Lett.~A {\bf 118}, 93 (1986).

\bibitem{Eshchenko_2002}
D.G.~Eshchenko, V.G.~Storchak, J.H.~Brewer, G.D.~ Morris, S.P.~Cottrell 
and S.F.J.~Cox,
Phys. Rev.~B {\bf 66}, 035105 (2002).

\bibitem{Hitti_2000}
B.~ Hitti, S.R.~Kreitzman, R.~Lichti, T.~Head, T.L.~Estle, D.~Wynne, 
E.E.~Haller, 
Physica B {\bf 289-290}, 554 (2000).

\bibitem{Chow_2005}
B.E.~Schultz, K.H.~Chow, B.~Hitti, R.F.~Kiefl, 
R.L.~Lichti, and S.F.J. Cox,
Phys. Rev. Lett. {\bf 95}, 086404 (2005).

\bibitem{Chow_1995}
K.H.~Chow, R.F.~Kiefl, W.A.~MacFarlane, J.W.~Schneider, 
D.W.~Cooke, M.~Leon, M.~Paciotti, T.L.~Estle, B.~Hitti, 
R.L.~Lichti, S.F.J.~Cox, C.~Schwab, E.A.~Davis, 
A.~Morrobel-Sosa, and L.~Zavieh,
Phys. Rev.~B {\bf 51}, 14762 (1995).

\bibitem{Holmes_1982}
D.E.~Holmes, R.T.~Chen, K.R.~Elliott, and Kirkpatrick,
Appl. Phys. Lett. {\bf 40}, 46 (1982).

\bibitem{Bourgoin_1996}
R.~Kiliulis, V.~Kazukauskas, and J.C.~Bourgoin,
J. Appl. Phys. {\bf 79}, 69511 (1996).

\bibitem{Martin_1979}	
A.~Mitonneau, A.~Mircea, G.M.~Martin, D.~Pons, 
Rev. Phys. Appl. {\bf 14}, 853 (1979).

\bibitem{Prinz_1983}
V.Ya.~Prinz and S.N.~Rechkunov,
Phys. Stat. Sol. (B) {\bf 118}, 159 (1983).  

%
%

\end{thebibliography}
\end{document}